\documentclass[aps,prl,amsmath,twocolumn,amssymb,titlepage,superscriptaddress,showpacs]{revtex4-1}
\pdfoutput=1
\usepackage[english]{babel}
\usepackage{graphicx}
\usepackage{transparent}
\usepackage{amsmath}
\usepackage{amssymb}
\usepackage{epstopdf}
\usepackage{amsfonts}
\usepackage{bm}
\usepackage{mathdots}
\usepackage{times}
\usepackage{mathtools}
\usepackage{stmaryrd}
\usepackage{hyperref}
\usepackage{tabularx}
\usepackage{hhline}

\usepackage{color}

\newcommand{\ave}[1]{\langle #1 \rangle}

\newcommand{\astcycl}{\mathrlap{\kern0.085em{\circlearrowright}}\ast}
\newcommand{\taucycl}{\mathrlap{\kern0.42em{\bullet}}\circlearrowright}

\begin{document}
\title{Nonequilibrium GW+EDMFT: Antiscreening and inverted populations from nonlocal correlations}
\author{Denis Gole\v z}
\affiliation{Department of Physics, University of Fribourg, 1700 Fribourg, Switzerland}
\author{Lewin Boehnke}
\affiliation{Department of Physics, University of Fribourg, 1700 Fribourg, Switzerland}
\author{Hugo U. R. Strand}
\affiliation{Department of Physics, University of Fribourg, 1700 Fribourg, Switzerland}
\author{Martin Eckstein}
\affiliation{Max-Planck Institute for the Structure and Dynamics of Matter, 22761 Hamburg, Germany}
\author{Philipp Werner}
\affiliation{Department of Physics, University of Fribourg, 1700 Fribourg, Switzerland}

\pacs{71.10.Fd,72.10.Di,05.70.Ln}

\begin{abstract}
We study the dynamics of screening in photo-doped Mott insulators with long-ranged interactions using a nonequilibrium implementation of the $GW$ plus extended dynamical mean field theory ($GW$+EDMFT) formalism. Our study demonstrates that the complex interplay of 
the injected carriers with 
bosonic degrees of freedom (charge fluctuations) can result in long-lived transient states with properties that are distinctly different from those of thermal equilibrium states. Systems with strong nonlocal interactions are found to exhibit a self-sustained population inversion of the doublons and holes. This population inversion leads to low-energy antiscreening which can be detected in time-resolved electron-energy loss spectra.  \end{abstract}
\maketitle

The development of time-resolved spectroscopic techniques provided important insights into the properties of complex materials \cite{giannetti2016,dalConte2012,dalconte2015,kemper2016,matveev2016}, where charge, spin, orbital and lattice degrees of freedom are intertwined.  A particularly exciting prospect is the nonequilibrium manipulation of material properties on electronic time scales, and the exploration of transient states that cannot be realized under equilibrium conditions. Prominent examples of this development are the laser-induced switching to a hidden state \cite{stojchevska2014} in 1T-TaS$_2$, and an apparent increase of the superconducting $T_c$ in phonon-driven cuprates and fulleride superconductors \cite{kaiser2014,mitrano2016}. 

Essential for the understanding of such experiments and phenomena is the ability to simulate relevant model systems using techniques that capture correlation effects in highly nonthermal states. Of particular importance is a proper description of the time-dependent screening processes, which determine the interaction parameters in such model Hamiltonians. The photo-induced change of screening was considered, e.g., as the cause of the collapse of the band gap in VO$_2$ \cite{wegkamp2014}, 
or for an enhancement of excitonic order in Ta$_2$NiSe$_5$ \cite{mor2016}. 
Moreover, screening originates from charge fluctuations, which,  similar to other bosonic modes like phonons \cite{allen87,werner2013,golez2012a,dorfner2015} or spin fluctuations \cite{golez2014,eckstein2016}, profoundly affect the relaxation pathway of the electronic distribution. As we will show in this paper, the fermionic dynamics and the bosonic screening modes are strongly coupled, 
so that their mutual interplay  can lead to long-lived transient states which are entirely different from those characterizing equilibrium phases. These non-thermal states, with partially inverted populations,  
thus provide an intriguing pathway to novel light-induced properties. 

A promising formalism to address these questions in strongly correlated solids is the combination of the $GW$ method and extended dynamical mean field theory ($GW$+EDMFT) \cite{biermann2003,werner2016a}. Hedin's $GW$ method \cite{hedin1965,aryasetiawan1998} is a weak coupling approach in which the self-energy is approximated by the product of the Green's function $G$ and the screened interaction $W$. It  captures nonlocal physics resulting from charge fluctuations, like screening, plasmonic collective modes and charge density waves. It however fails to describe strong correlation effects, like the Mott metal-insulator transition, which in turn are well described by the non-perturbative dynamical mean field theory (DMFT)  \cite{georges1996} and extended DMFT (EDMFT) \cite{sun2002}. $GW$+EDMFT is a fully diagrammatic approach, which allows a self-consistent calculation of the screened interaction and its effect on the electronic properties in systems with long-ranged Coulomb interactions, and, in combination with a $GW$-based ab-initio simulation, a parameter-free simulation of weakly and strongly correlated  materials.  The recent equilibrium application of $GW$+EDMFT to model systems \cite{ayral2013,huang2014,ayral2017} and real materials \cite{boehnke2016} demonstrated the importance of dynamical screening effects originating from nonlocal interactions, e.g., for the proper interpretation of spectral features such as Hubbard bands and plasmon satellites. Here, we develop the nonequilibrium extension of the $GW$+EDMFT formalism and use it to study the effect of nonlocal interactions on the transient states and the relaxation dynamics of photoexcited carriers in Mott insulators. 

As a simple but generic system with inter-site interactions we consider the single-band $U$-$V$ Hubbard model on the two-dimensional square lattice,
\begin{multline}
H(t)=-J\sum_{\langle ij \rangle\sigma}(e^{i\phi_{ij}(t)}c_{i\sigma}^{\dagger}c_{j\sigma}+h.c.)
-\mu \sum_{i}n_i \\
+\sum_{i} U (n_{i\uparrow} \!-\! \tfrac{1}{2})(n_{i\downarrow}
\!-\! \tfrac{1}{2}) + \sum_{\langle ij\rangle} V(n_{i}\!-\!1)(n_{j}\!-\!1),
\label{Eq.:Hubbard-UV}
\end{multline}
where $c_{i\sigma}$ is the annihilation operator of a fermion with spin $\sigma$ on lattice site $i$, $n_{i}=n_{i\uparrow}+n_{i\downarrow}$, 
$\mu$ is the chemical potential, $U$ the on-site interaction, and $V$ the interaction between electrons on neighboring sites \footnote{The extension to longer ranged interactions is straightforward \cite{huang2014} and does not lead to qualitatively new physics.}.
The hopping integral $Je^{i\phi_{ij}(t)}$ (restricted to nearest neighbors) has a time-dependent Peierls phase $\phi_{ij}(t)=\int^t_0 d\bar t\, \vec{E}(\bar t)(\vec{r}_i-\vec{r}_i)$ originating from an in-plane electric field $\vec{E}(t)$. In the following we will use the hopping amplitude $J\equiv 1$ as the unit of energy, and rewrite the interaction as $\frac{1}{2}\sum_{ij}v_{ij}\tilde n_{i}\tilde n_{j}$, where $\tilde n=n-1$ is  the density fluctuation operator, and  $v_{ij}=U\delta_{ij}+V\delta_{\langle ij\rangle}$.

The dynamics of the system is described in terms of the momentum-dependent electron Green's function $G_k(t,t')$
$= -i \langle T_\mathcal{C} c_{k}(t)  c_{k}^\dagger (t')\rangle$, and the charge correlation function $\chi_{q}(t,t')=-i\ave{  T_\mathcal{C} \tilde n_q(t) \tilde n_{-q}(t')}$, which determines the (inverse) dielectric function $\varepsilon^{-1}_q=1+v_q\ast \chi_q$, and the screened interaction $W_q=\varepsilon^{-1}_q \ast v_q$, where $v_q$ is the Fourier transform of $v_{ij}$. In nonequilibrium, all quantities depend on two time arguments, or equivalently on time and frequency, and the \mbox{$\ast$-product} denotes convolution in time \footnote{For a precise definition of time ordered correlation functions on the Keldysh contour, see the Supplemental Material\cite{supp}}.

To solve the extended Hubbard model in Eq.~(\ref{Eq.:Hubbard-UV}) we resort to the $GW$+EDMFT approximation \cite{biermann2003}, which can be derived using the 
Almbladh functional \cite{almbladh1999}. Nonlocal self-energy contributions for electrons and bosonic charge fluctuations are treated within the lowest order expansion of the functional (the $GW$ formalism), while the local contributions are included to all orders, by solving an auxiliary Anderson-Holstein impurity model with a self-consistently determined bosonic and fermionic bath.  As a Green's function based formalism, $GW$+EDMFT is not restricted to equilibrium or quasi-static problems, but can handle highly excited states. The
derivation of the nonequilibrium formalism within the Keldysh framework is analogous to the equilibrium version~\cite{ayral2013,huegel2016}, and is presented in the Supplemental Material~\cite{supp}.

\begin{figure}[t]
\includegraphics{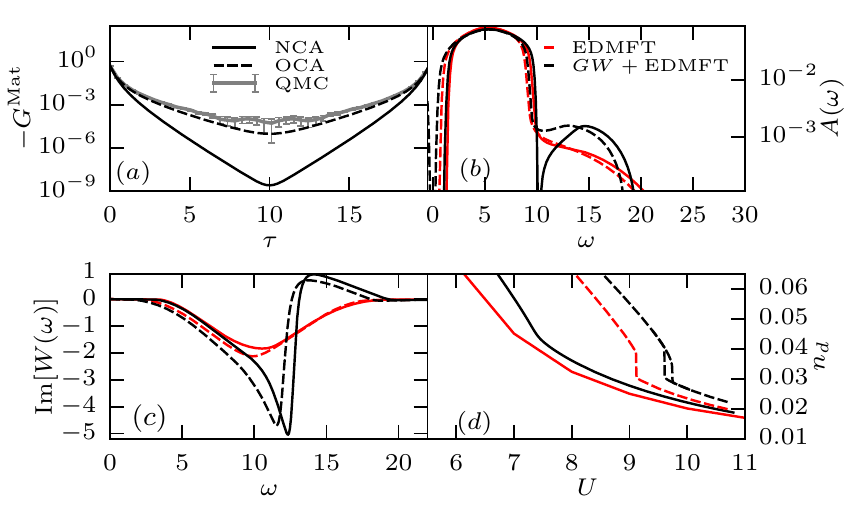}
\caption{Equilibrium results for $U=10.5$, $V=1.5$ and $\beta=20$. (a) Comparison of the Matsubara time component of the Green's function $G^{\text{Mat}}(\tau)$ obtained from NCA, OCA and numerically exact Monte Carlo for $U=10.5$. (b) Spectral functions obtained from different approximations. 
Full (dashed) lines correspond to the NCA (OCA) solution. 
(c) Imaginary part of the screened interaction $W(\omega)$ obtained from different approximations. (d) Double occupation $n_d$ near the metal insulator transition or crossover. A coexistence region exists in the OCA approximation. 
}
\label{Fig:Comparison}
\end{figure}

While powerful and numerically exact methods \cite{werner2010} exist for the solution of the $GW$+EDMFT equations in equilibrium, the application to nonequilibrium problems requires additional approximations at the level of the impurity solver. Since our goal is to study photo-doped Mott insulators, we use a perturbative solver that combines a self-consistent hybridization expansion (at first (second) order known as the non-crossing (one-crossing) approximation NCA (OCA) \cite{eckstein2010,grewe1981,coleman1984}) with a weak-coupling expansion in the retarded density-density interactions. For technical aspects of the implementation, see Ref.~\onlinecite{golez2015}. As a benchmark, we show in Fig.~\ref{Fig:Comparison}(a) a comparison of the Matsubara component of the Green's functions $G^{\text{Mat}}(\tau)$ for $U=10.5$, $V=1.5$ and inverse temperature $\beta=20$. (In the following local (nonlocal) correlators are distingushed by the absence (presence) of a subscript momentum label.) The NCA is found to overestimate the insulating nature of the solution \cite{pruschke1989,haule2001}, as seen from $G^{\textrm{Mat}}(\beta/2)$, which can be taken as a measure for the spectral weight at the Fermi level. While this is a known artefact of the NCA~\cite{pruschke1989,haule2001}, the OCA substantially improves the accuracy of the solution compared to numerically exact Monte Carlo results~\cite{werner2010}. Furthermore,  the finite-temperature metal-insulator transition is a crossover in the NCA description and becomes first order in the OCA solution, see
Fig.~\ref{Fig:Comparison}(d). In the Mott phase, which we study here, NCA and OCA however yield qualitatively similar results, and we will resort to the numerically more tractable NCA in the following.

In the spectral function, shown in Fig.~\ref{Fig:Comparison}(b), the additional nonlocal $GW$ self-energy contributions in $GW$+EDMFT strongly enhance the plasmonic sideband at $\omega \approx \tfrac 32 U$ and result in a slight reduction of the gap size compared to EDMFT.  
The inclusion of the nonlocal $GW$ diagrams in the $GW$+EDMFT approximation leads to a more metallic solution, since nonlocal correlations (in particular the nonlocal Fock term \cite{ayral2017}) enhance the effective bandwidth. Also the local (momentum averaged) screened interaction $W$ is modified by the inclusion of the nonlocal polarization [Fig.~\ref{Fig:Comparison}(c)]. A noticeable feature is the strong enhancement of the plasmonic peak at $\omega\approx12$ in comparison to EDMFT. A drawback of our approximate solver is evident at energies above the plasmon peak, where $\textrm{Im}[W]$ exhibits positive spectral weight, which is unphysical in thermal equilibrium. This problem arises because the NCA and OCA self-energies and polarizations are approximate strong-coupling solutions, which miss some of the local $GW$ diagrams. Numerically we found that these artefacts are most pronounced deep in the Mott phase, while close to the MIT transition and in the correlated metal $\textrm{Im}[W(\omega)]$ exhibits the expected analytical properties. Since the unphysical spectral weight appears only at very high energies, we believe that it is not crucial for the following discussion, which focuses on the low-energy screening properties of photodoped systems.

\begin{figure}[t]
\includegraphics[width=1.\linewidth]{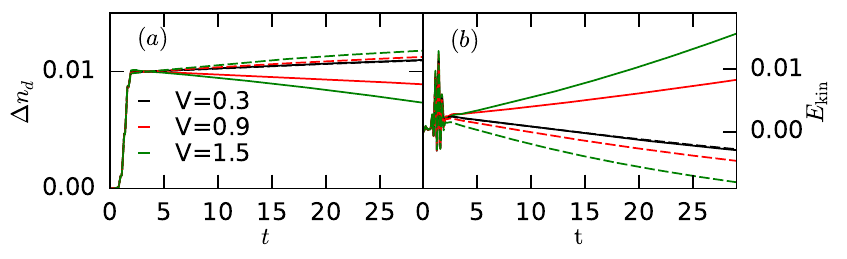}
\caption{Time evolution of the double occupancy $n_d$ (a) and kinetic energy (b) after the photo excitation in EDMFT (dashed) and $GW$+EDMFT (full lines) for different nonlocal interactions $V$ at fixed density $\Delta  n_d=0.01$ of photo-excited carriers after the pulse. The local interaction is $U=10.5$. 
}
\label{Fig:Double}
\end{figure}

We now turn to the effect of nonlocal fluctuations on the relaxation dynamics after an electric field excitation. By applying a short pulse 
$E(t)=E_0 e^{-4.6(t-t_0)^2/t_0^2} \sin(\omega (t-t_0))$
with frequency $\omega=U$ 
and appropriately tuned amplitude $E_0$ a certain density of 
holon-doublon pairs is created. 
The width of the pulse  $t_0=2\pi n/\omega$  is chosen such that the envelope accommodates $n=2$ electric field cycles. Deep in the Mott phase, the recombination of the holons and doublons after photo-excitation is strongly suppressed \cite{sensarma2010,eckstein2010,zala2013}. The photoexcited doublons can however relax within the upper Hubbard band, which manifests itself in the evolution of the kinetic energy. If the gap is small compared to the width of the Hubbard bands, the thermalization process, which involves impact ionization \cite{werner2014}, leads to an {\it increase} in the number of doublons $n_d$,  
see the EDMFT results (dashed lines) in Fig.~\ref{Fig:Double}(a).  As already discussed in Ref.~\cite{golez2015} the inclusion of the nonlocal interactions on the EDMFT level decreases the relaxation times, due to the coupling to bosonic excitations (collective charge fluctuations). This picture remains valid if we include nonlocal self-energy and polarization effects in $GW$+EDMFT, but only if the nearest-neighbor interaction $V$ is small ($V \lesssim 0.5$).  
For larger values of $V$  (but still smaller than the critical value for the  
charge 
order transition), the double occupancy starts to {\it decrease}, which indicates that doublon-holon recombination occurs in the system, see solid lines in Fig.~\ref{Fig:Double}(a). 
Furthermore, the kinetic energy {\it increases} during the relaxation process, illustrated in Fig.~\ref{Fig:Double}(b), which is also intriguingly different from the behavior reported in previous photodoping studies~\cite{werner2014,werner2013,golez2015}. 

\begin{figure}[t]
\includegraphics{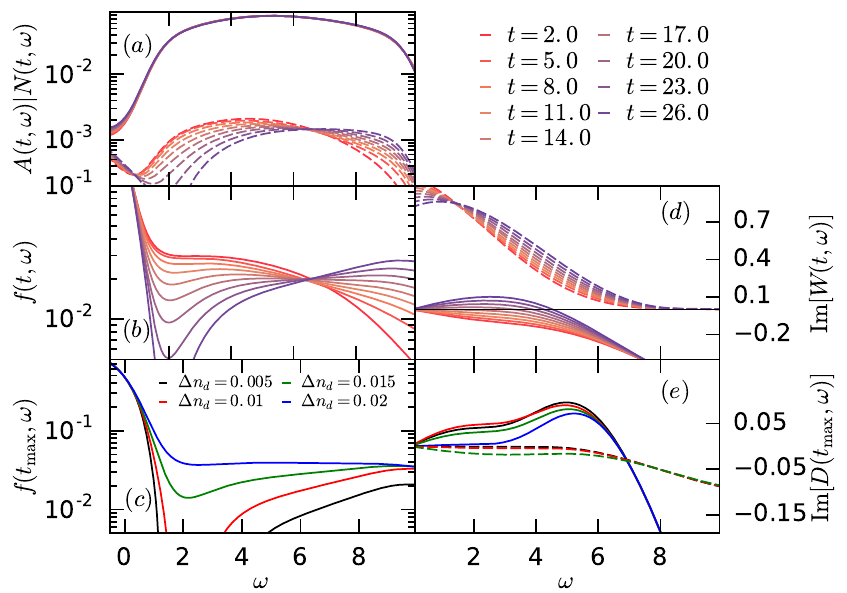}
\caption{(a) Time evolution of the spectral function (full) and occupation (dashed) after the electric field excitation. (b) The distribution function illustrates the evolution into the self-sustained inverted population state. (c) Distribution functions for $t_\text{max}=24$ and different excitation strengths. (d) Time evolution of the screened interaction $W^R(t,\omega)$ (solid) and its lesser component (boson occupancy, dashed) $W^<(t,\omega)$ 
in the inverted population regime. (e) Imaginary part of the impurity effective interaction $\text{Im}[D^{R}(t_\text{max},\omega)]$ in EDMFT (dashed) and $GW$+EDMFT (full) for different excitations strengths. The pulse frequency is $\omega=U=10.5$, the pulse amplitude is $E_0=2$ (except in panel (c)), and $V=1.5$.}
\label{Fig:Distribution}
\end{figure}

In order to gain further insight into this intermediate $V$ regime we calculate the time and frequency-resolved spectral function of the system.  After the pulse excitation of the system the spectral function $A(t,\omega) = -\frac{1}{\pi}\text{Im}[G^R(t,\omega)]$ remains almost unchanged, while the occupied density of states $N(t,\omega)=\textrm{Im}[G^<(t,\omega)]/2\pi i$ shows an increase of  roughly 1\%  in the occupancy of the upper Hubbard band \footnote{For all two-time quantities $O(t,t')$ we define the partial Fourier-transform as $O(t,\omega)=\int_{t}^{t+t_\text{cut}} dt' e^{\mathrm{i} \omega (t'-t) }O(t',t)$, with $t_{\text{cut}}=7$, which allows us to compute the spectrum with a time-independent resolution up to relatively long times. We checked that this choice of cut-off does not qualitatively affect the dynamics of the spectral properties under consideration.}, see Fig.~\ref{Fig:Distribution}(a).  
In agreement with the evolution of the kinetic energy, we observe a shift of the excited doublons toward {\it higher energies}, 
in contrast to previous DMFT and EDMFT studies \cite{werner2013,werner2014,golez2015} that consistently showed a relaxation of 
doublons to the lower edge of the upper Hubbard band. This $GW$+EDMFT evolution eventually results in a population inversion, as illustrated by the distribution function 
$f(t,\omega)=-2\text{Im}[G^<(t,\omega)]/\text{Im}[G^R(t,\omega)]$  
shown in Fig.~\ref{Fig:Distribution}(b).
We note again that this behavior is observed 
only for sufficiently large nonlocal interaction $V$.

\begin{figure*}[ht]
\includegraphics{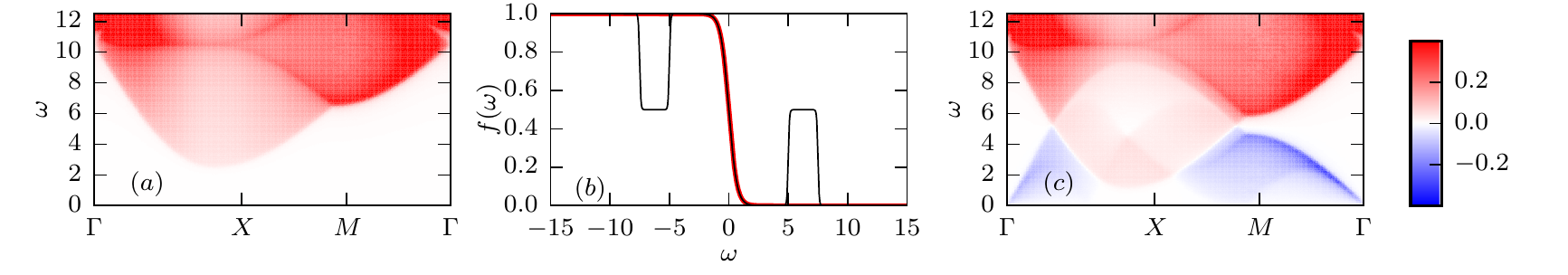}
\includegraphics{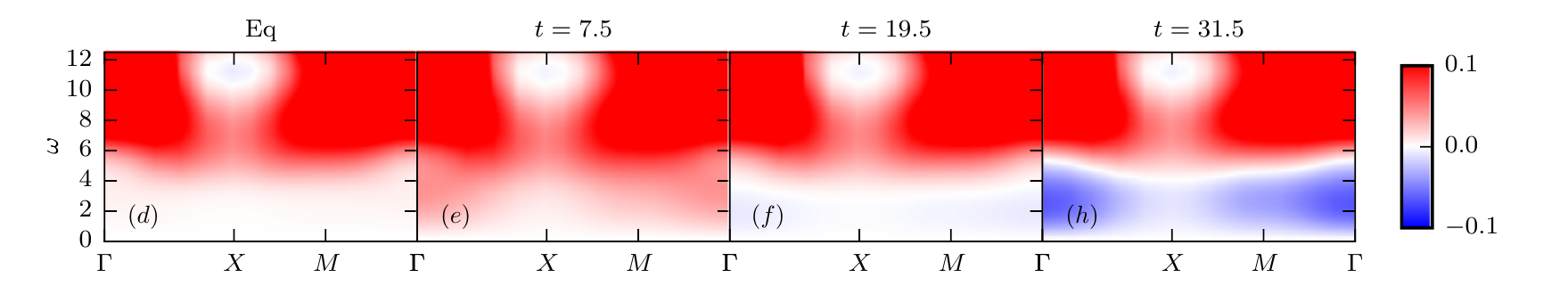}
\caption{Top panels: Imaginary part of the inverse dielectric function
$-\text{Im}[\varepsilon_q^{-1}(\omega)]$ obtained in the Hubbard I approximation for the thermal (panel a) and non-thermal (panel c) distribution functions shown in panel b. Bottom panels: $GW$+EDMFT results for $-\text{Im}[\varepsilon_q^{-1}(t,\omega)]$ in equilibrium (panel d) and in the photoexcited system at indicated time delays (panels e-h). The pulse frequency is $\omega$ = $U$ = 10.5, the pulse amplitude is $E_0=2$, and $V=1.5$.}
\label{Fig:EELS}
\end{figure*}

The efficient recombination of doublon-hole pairs and the population inversion within $GW$+EDMFT
can be understood by considering the two-particle properties, namely the screened interaction $W$ and the charge susceptibility $\chi_q.$  The time evolution of the local component of the screened interaction $W^{R,<}(t,\omega)$ for $U=10.5,V=1.5$
is shown in Fig.~\ref{Fig:Distribution}(d).
In agreement with the previous EDMFT results, low energy screening channels appear as a consequence of photo-doping \cite{golez2015}. The main difference in $GW$+EDMFT is that the imaginary part of $W^R(t,\omega)$ changes sign as the system evolves into the population-inverted state. Since EDMFT and $GW$+EDMFT differ in the inclusion of nonlocal fluctuations we can qualitatively understand these results by evaluating the nonlocal charge susceptibility through the particle-hole bubble contribution to the polarization. In the stationary case the latter can be written as $\chi_q^\mathrm{R}=\Pi_q^\mathrm{R}[1-v_q\Pi_q^\mathrm{R}]^{-1}$ where the polarization $\Pi_q^R$ is given by
\begin{equation}
  \Pi_{q}^\mathrm{R}(\omega)=
  \sum_{k,\omega_1,\omega_2} A_k(\omega_1) A_{k-q}(\omega_2) \frac{f(\omega_1)- f(\omega_2)}{\omega-(\omega_1-\omega_2)}, 
\label{Eq.:chikw}
\end{equation}
which, in the case of well defined quasi-particles and thermal distributions $f$, reduces to the Lindhard formula.
By exciting doublon-hole pairs in a Mott insulator, 
we temporarily create an inverted population in some energy range. Changing the Fermi-Dirac distribution function $f(\omega)$ in 
Eq.~\eqref{Eq.:chikw} to a partially inverted distribution function $\tilde f(\omega)$, we can change the sign of the numerator in $\chi_q^\mathrm{R}$
within a certain energy range. 
To illustrate this idea we evaluate $\chi_q^\mathrm{R}(\omega)$ using the Hubbard I approximation, where the lattice self-energy is approximated by the atomic limit self-energy. The resulting inverse dielectric function is shown in Fig.~\ref{Fig:EELS}(a) where $\text{Im}[\varepsilon^{-1}_q(t,\omega)]=v_q \text{Im}[\chi_q^\mathrm{R}(t,\omega)]$. This leads to maximum spectral weight at the $\Gamma$-point and $\omega \approx U,$ which corresponds to charge excitations across the Mott gap. The lowest (highest) energies $U\pm W$ for which the imaginary part of the susceptibility $\text{Im}[\chi_q^\mathrm{R}(\omega)]$ has non-zero weight are at the $X$-point [$q=(\pi,\pi)$] \footnote{An further increased nonlocal interaction strength $V$ would  eventually lead to the condensation of the bosonic modes at $(\pi,\pi)$ and the formation of charge order.}. In the case of the inverted population, see Fig.~\ref{Fig:EELS}(c), the numerator in Eq.~(\ref{Eq.:chikw}) becomes negative at  
frequencies  
corresponding to the energy width of the inverted regions, which leads to a negative spectral weight $-v_q\text{Im}[\chi_{q}^R(\omega)]<0$. These considerations show that the inclusion of nonlocal dynamical screening via the polarization bubble in $GW$+EDMFT is crucial for the appearance of the anti-screening phenomenon.

In contrast to the fermionic case, negative spectral weight in a steady state bosonic spectral function is not unphysical. The simplest example is a free oscillator, whose  frequency suddenly turns unstable ($\omega_0<0$). Although there is no stable thermal equilibrium for $\omega_0<0$, the transient state remains well-defined, and its negative spectral weight  reflects the possibility to increase fluctuations by emitting energy to the environment.
The change of the sign of $\text{Im}[\chi_q(\omega)]$ in the photo-doped Mott insulator thus indicates a negative attenuation of charge fluctuations, which enable the system to emit low energy bosons to gain energy in the single particle sector. This also explains the unusual increase of the kinetic energy within the upper Hubbard band and the population inversion.  A similar change in the sign of the susceptibilities was previously observed in models which are driven by (time-periodic) external fields \cite{tsuji2009,tsujiPhd}.  The intriguing observation in the present case is that the inverted population of the electronic states and the negative charge susceptibility mutually support each other (because the softening of charge fluctuations is caused by the change of the fermionic distribution), so that the peculiar state is self-sustained and stable as long as doublon-hole recombination processes inject energy into the bosonic subsystem. 

A related population inversion was recently discussed in a study of Hirsch's dynamic Hubbard model \cite{hirsch2001}, although at unusually strong electron-phonon couplings. In the present case, the relevant strength $\lambda$ of the electron-boson coupling can be estimated from the density of states $D(t,\omega)$ of the bosonic modes in the auxiliary Anderson-Holstein impurity model
(i.e., the boson-mediated density-density interaction interaction) as $\lambda = \int  d\omega \sqrt{ |\text{Im}D(\omega)|\omega}$~\cite{ayral2013,golez2015}. As shown in Fig.~\ref{Fig:Distribution}(e), in $GW$+EDMFT, $\text{Im}|D(t,\omega)|$ features a pronounced peak at the energy of the gap size $\omega\approx 6,$ which corresponds to a very strong electron-boson coupling ($\lambda\approx 1.9$ for the largest value of $E$ plotted in the figure, if the integration range is chosen as $0\le \omega\le 8$). 

Experimental probes which could be used to detect the peculiar charge fluctuation region are electron energy-loss spectroscopy (EELS) \cite{kogar2014,ibach2013} and optical conductivity measurements \cite{giannetti2016}. The optical conductivity measures the frequency dependent optical constant near the $\Gamma$-point \cite{eckstein2008,zala2014}, while the EELS signal $-\text{Im}[v_q\varepsilon_q^{-1}(t,\omega)]=-v_q^2 \text{Im}[\chi_q^\mathrm{R}(t,\omega)]$  measures the difference between the dielectric loss and gain (in equilibrium and at low temperatures, there is only loss). The generalization of EELS to the non-equilibrium situation, along the lines of the derivation of the time-dependent photo-emission formula \cite{eckstein2008,freericks09,randi2017}, is presented in the Supplementary Material \cite{supp}.
The closely related inverse dielectric constant
$\varepsilon_q^{-1}(t,\omega)$ shows a similar structure in $GW$+EDMFT as in the Hubbard I approximation, see Fig.~\ref{Fig:EELS}(a) \& (d). In particular, there is a pronounced maximum at the $\Gamma$-point at $\omega\approx U$ and dispersive bands with a minimal energy around the $X$-point. Immediately after the excitation the weight in the sub-gap region is increased in agreement with previous EDMFT results \cite{golez2015}, see Fig.~\ref{Fig:EELS}(e). The initial increase in the screening in the sub-gap region 
however gives way to a negative 
spectrum as the inverted doublon population is formed [Fig.~\ref{Fig:EELS}(f)-(h)], and the bosonic degrees of freedom 
also evolve into an inverted state. In this situation the energy gain for the probe electron at a certain energy in the EELS experiment is larger than the loss.

In conclusion, the nonequilibrium $GW$+EDMFT simulation revealed a self-sustained and long-lived transient population inversion as a result of the nontrivial energy exchange between doublons, holons and charge fluctuations. The existence of such a state provides an intriguing path to stabilize different types of light-induced  order, which will be the subject of future investigations. Apart from these insights into the nonequilibrium properties of systems with nonlocal Coulomb interactions, our work represents an important step in the development of ab-initio simulation approaches for correlated systems in nonequilibrium states. The $GW$+EDMFT method implemented here features a fully consistent treatment of correlation and screening effects, and can in principle be combined with material-specific input from ab-initio $GW$ calculations within a multi-tier approach analogous to the scheme recently demonstrated for equilibrium systems in Ref.~\onlinecite{boehnke2016}.

{\it Acknowledgements}
The calculations have been performed on the Beo04 cluster at the University of Fribourg. DG, LB, HS and PW have been supported by ERC starting grant No. 278023 and the SNSF through NCCR MARVEL and Grant No. 200021-165539. ME acknowledges support by the Deutsche Forschungsgemeinschaft within the Sonderforschungsbereich 925 (project B4).

\end{document}


\title{Supplementary Material}
\author{Denis Gole\v z}
\affiliation{Department of Physics, University of Fribourg, 1700 Fribourg, Switzerland}
\author{Lewin Boehnke}
\affiliation{Department of Physics, University of Fribourg, 1700 Fribourg, Switzerland}
\author{Hugo U. R. Strand}
\affiliation{Department of Physics, University of Fribourg, 1700 Fribourg, Switzerland}
\author{Martin Eckstein}
\affiliation{Max Planck Research Department for Structural Dynamics, University of Hamburg-CFEL, 22761 Hamburg, Germany}
\author{Philipp Werner}
\affiliation{Department of Physics, University of Fribourg, 1700 Fribourg, Switzerland}
\maketitle
\label{functional}

\section{Functional derivation of GW+EDMFT}
We consider the action formalism, in which consistent approximations can be derived from a common functional \cite{kadanoff1962} and in which the generalization to real time dynamics is easily obtained by the Baym-Kadanoff formalism. The general strategy is to construct a Baym-Kadanoff functional $\Gamma$ by a Legendre transform of the free energy $\Omega$. The free energy $\Omega$ is a functional of the bare propagator $G_0$ and the interaction $v$, $\Omega \equiv \Omega[G_0, v]$. By the Legendre transform the functional dependence is changed, making $\Gamma$ (except for the Hartree part $\Gamma^\mathrm{H}$) a functional of the (interacting) single particle Green's function $G_{ij}(t,t')=-i \ave{T_\mathcal{C}c_i(t) c_j^{*}(t')}$ and the screened interaction $W_{ij}(t,t')$, $\Gamma \equiv \Gamma[G, W]$. At the physical $G$ and $W$ the Legendre transform guarantees that $\Gamma$ is stationary and takes the value of the free energy $\Omega$. Via a suitable Hubbard-Stratonovich (HS) transformation \cite{ayral2013,chitra2001} the screened interaction $W_{ij}(t,t')$ can be shown to be related to the charge susceptibility $\chi_{ij}(t,t')=-i\ave{T_\mathcal{C}\tilde n_i(t) \tilde n_j(t')}$ through the integral relation
\begin{equation}
  W=v+v*\chi*v ,
\end{equation}
where the $*$-product denotes a convolution on the Baym-Kadanoff L-shaped time-contour and a sum over (adjacent) real space indices. The HS transformation yields a scalar electron-boson vertex $\alpha$ equal to unity, $\alpha = 1$, however, in the functional treatment it is convenient to also consider the $\alpha = 0$ case, where the electron and boson systems are decoupled.

The 
Baym-Kadanoff functional $\Gamma$ can be written as
\beq{
  \Gamma_{\alpha=1}[G,W]=\Gamma_{\alpha=0}[G,W]+\Gamma^\mathrm{H}[G,v]+\Psi[G,W],
}
where $\Gamma_{\alpha=0}$ is the decoupled part of $\Gamma$,
\bsplit{
  \Gamma_{\alpha=0}=&\Tr[\ln(-G)]-\Tr[G_0^{-1}*G]\\
  &-\frac{1}{2}\Tr[\ln(W)]+\frac{1}{2}\Tr[v^{-1}*W].
}
The Hartree contribution to the functional, $\Gamma^\mathrm{H}=-2\tfrac{i}{2} \Tr[ G_{ii}(t,t^+)v_{ij}G_{jj}(t,t^+) ]$, only depends on the bare interaction and has to be treated separately. The factor of 2 originates from the sum over spin degrees of freedom. The remaining many-body complexity of the system is now captured by the Almbladh functional $\Psi \equiv \Psi[G, W]$, comprising all possible two-particle irreducible diagrams built with the $G$ and $W$ propagators \cite{almbladh1999}.
The physical solution corresponds to the stationary points of the Baym-Kadanoff functional $\Gamma$ with respect to $G$ and $W$, namely $\frac{\delta\Gamma}{\delta G}=0$ and $\frac{\delta\Gamma}{\delta W}=0$, which yields the Dyson equations for $G$ and $W$:
\beq{
  G^{-1}=G_0^{-1}-\Sigma^\mathrm{H}-\Sigma^{\mathrm{xc}}, \qquad W^{-1}=v^{-1}-\Pi,
}
wherein $\Sigma^\mathrm{H}=\delta_G \Gamma^\mathrm{H}[G,v]$ is the Hartree self energy, and the exchange-correlation self energy $\Sigma^{\mathrm{xc}}$ and the polarization $\Pi$ are obtained from variations of the Almbladh functional $\Psi$,
\beq{
  \Sigma^{\mathrm{xc}}=\frac{\delta \Psi}{\delta G},\qquad
  \Pi=-2\frac{\delta \Psi}{\delta W}.
}

Evaluating all diagrams in $\Psi$ is not a tractable problem, and therefore we seek approximations that keep only a subset of the diagrams.
$GW$ is such an approximation, which retains only the lowest order contribution in the electron-boson coupling $\alpha$ (apart from the Hartree term, which is treated separately). It corresponds to the following approximation of the Almbladh functional: 
\beq{
\Psi \approx \Psi_{GW} \equiv \frac{i}{2} \Tr[ G_{ij}(t,t') W_{ij}(t, t') G_{ji}(t', t) ]. 
}
The resulting approximations for the self energy, $\Sigma^{GW}_{ij}(t,t')=i G_{ij}(t,t')W_{ij}(t,t')$, and polarization, $\Pi^{GW}_{ij}(t,t')=-i G(t,t')_{ij}G(t',t)_{ji}$, provide a decent description of weakly correlated systems and capture charge fluctuation driven nonlocal physics, like screening, plasmonic collective modes, and charge density waves. However, as a weak coupling expansion, it fails to describe effects of strong correlations, like Mott's metal-insulator transition.

An approximation that captures these latter phenomena is extended dynamical mean field theory (EDMFT), which corresponds to the following local approximation 
of the Almbladh functional:
\beq{
  \Psi \equiv \Psi[G_{ij},W_{ij}] \approx \Psi[G_{ii},W_{ii}] \equiv \Psi_{\mathrm{EDMFT}}.
}
Note that this is a highly non-perturbative approximation that accounts for all diagrams, which contain only local propagators.

In order to capture both the effect of strong interactions and nonlocal physics we can combine the two functionals \cite{biermann2003},
by 
supplementing the 
local diagrams in $\Psi_\mathrm{EDMFT}$ with all nonlocal GW diagrams,
\begin{multline}
  \Psi \approx \Psi_{GW+\mathrm{EDMFT}} \equiv
  \Psi_\mathrm{EDMFT}[G_{ii},W_{ii}] \\
  + \Psi_{GW}[G_{ij},W_{ij}] -\Psi_{GW}[G_{ii},W_{ii}],\label{Eq.:Almbladh_E+GW}
\end{multline}
arriving at the $GW$+EDMFT approximation of the Almbladh functional $\Psi_{GW+\mathrm{EDMFT}}$.

The basic insight of the EDMFT approach is the observation that there exists a solvable many-body problem, whose Almbladh functional is given by $\Psi_\mathrm{EDMFT}[G_{ii}, W_{ii}]$, namely an effective impurity problem (for the simpler DMFT case see Ref.\ \onlinecite{huegel2016}). In order to evaluate the self-energy contributions from $\Psi_\mathrm{EDMFT}[G_{ii}, W_{ii}]$ we want to construct this impurity system
so that it has the local Green's function $G_{ii}$ and screened interaction $W_{ii}$ of the lattice problem. This is achieved by deriving constraining equations for the \textit{a priori} unknown impurity Weiss field $\mathcal{G}$ and effective impurity interaction $\mathcal{U}$. The Baym-Kadanoff functional $\Gamma'$ of the 
impurity problem can be expressed as
\beq{
  \Gamma'=\Gamma_0' +\Gamma^\mathrm{H}{}' +\Psi_\mathrm{EDMFT}[G_{ii},W_{ii}],
}
where the decoupled contribution is 
\bsplit{
\Gamma_0'=&\Tr[\ln(-G_{ii})]-\Tr[\mathcal{G}^{-1}*G_{ii}]\\
&-\frac{1}{2}\Tr[\ln(W_{ii})]+\frac{1}{2}\Tr[\mathcal{U}^{-1}*W_{ii}],
}
and the Hartree contribution is given by
\beq{
  \Gamma^\mathrm{H}{}'=-2\tfrac{i}{2} \Tr[ G_{ii}(t,t^+) \mathcal{U}(t,t') G_{ii}(t',t'^+) ]. 
}
Demanding that the lattice system functional $\Gamma$ and the impurity system functional $\Gamma'$ have the same local $G_{ii}$ and $W_{ii}$ causes both functionals to be stationary ($\delta_{G_{ii}}\Gamma'=0=\delta_{G_{ij}}\Gamma$ and $\delta_{W_{ii}}\Gamma'=0=\delta_{W_{ij}}\Gamma$ or, for convenience~\cite{huegel2016}, $\delta_{G_{ii}}(\Gamma-\Gamma')=\delta_{G_{ii}}\Gamma_{GW+\mathrm{EDMFT}}=0$, $\delta_{W_{ii}}(\Gamma-\Gamma')=\delta_{W_{ii}}\Gamma_{GW+\mathrm{EDMFT}}=0$), which yields the conditions for the auxiliary 
quantities 
$\mathcal{G}$ and $\mathcal{U}$ that will have to be satisfied by the selfconsistent solution.
We have the following explicit expressions:
\begin{align}
&\frac{\delta \Gamma_{GW+\mathrm{EDMFT}}}{\delta G_{ii}}=-[G_0^{-1}]_{ii}+[G^{-1}]_{ii}+\mathcal{G}^{-1}-[G_{ii}]^{-1}\nonumber\\
&\hspace{30mm}+\Sigma^{\mathrm{H}}-\Sigma^{\mathrm{H}}{}'=0,\nonumber\\
&\Rightarrow \mathcal{G}^{-1}=[G_{ii}]^{-1}+\Sigma^{\mathrm{xc}}_{ii}+\Sigma^{\mathrm{H}}{}',\label{eqG}
\end{align}
\begin{align}
&\frac{\delta \Gamma_{GW+\mathrm{EDMFT}}}{\delta W_{ii}}=[v^{-1}]_{ii}-[W^{-1}]_{ii}-\left(\mathcal{U}^{-1}-[W_{ii}]^{-1}\right)=0,\nonumber\\
&\Rightarrow \mathcal{U}^{-1}=[W_{ii}]^{-1} + \Pi_{ii}.\label{eqW}
\end{align}
Equations (\ref{eqG}) and (\ref{eqW}) constitute the self-consistency condition for the $GW$+EDMFT approach.

The final Dyson equations of the lattice system in the $GW$+EDMFT approximation thus read
\begin{gather}
  [G_0^{-1}]_{ij} \!-\! [G^{-1}]_{ij} =  \Sigma^H_{ij} + \delta_{ij}\Sigma^{\textrm{xc}}_{ii} + (1 \!-\! \delta_{ij})\Sigma^{GW}_{ij},
  \\
  [v^{-1}]_{ij} - [W^{-1}]_{ij}  = \delta_{ij}\Pi_{ii} + (1 - \delta_{ij})\Pi^{GW}_{ij},  
\end{gather}
where $\Sigma^{\textrm{xc}}_{ii}$ and  $\Pi_{ii}$ are obtained from Eqs.~(\ref{eqG}) and (\ref{eqW}).

In the actual implementation the self-energy contributions are grouped slightly differently to take advantage of the Hartree and Fock self-energy contribution $\Sigma^\mathrm{HF}$ being instantaneous
\bsplit{
  \Sigma_{k}^{GW+\mathrm{EDMFT}}=&\Sigma^{\mathrm{HF}}_{k}+\Sigma^{GW_c,\mathrm{nl}}_{k}+\Sigma^\mathrm{xc}_{ii}\\
  =&[\Sigma^\mathrm{HF}_{k}+\Sigma^\mathrm{H}{}']+\Sigma^{GW_c,\mathrm{nl}}_{k}+\\
  &[\Sigma^\mathrm{xc}_{ii}-\Sigma^\mathrm{H}{}'],
}
where $\Sigma^{GW_c,\mathrm{nl}}=\Sigma_k^{GW_c}-\sum_k\Sigma_k^{GW_c}$ is the nonlocal part and $\Sigma^{GW_c}_{ij}(t,t')=iG_{ij}(t,t')(W_{ij}-v_{ij})(t,t')$ is the $GW$ self-energy with the Fock term removed.

The last bracket on the right hand side follows naturally when the 
combined weak-coupling and hybridization expansion
is performed in terms of the density fluctuations $\tilde{n}=n-\ave{n}$ instead of the density \cite{eckstein2010,golez2015}:
\bsplit{
&\frac{1}{2}\int\int dt dt' \tilde n(t) \mathcal{U}(t,t')\tilde n(t')=\frac{1}{2}\int\int dt dt' n(t) \mathcal{U}(t,t')  n(t') \\
&-\ave{n} \int dt dt'  \mathcal{U}(t,t')  n(t')+ \ave{n}\ave{n} \int dt dt'  \mathcal{U}(t,t').
}
Then, the second term on the r.h.s.\ cancels the impurity Hartree contribution $\Sigma^{\mathrm{H}'}(t,t')=\delta_{\mathcal{C}}(t,t')\ave{n} \int d\bar t \mathcal{U}(t,\bar t)$, where $\delta_{\mathcal{C}}$ marks the Delta function on the Baym-Kadanoff contour.

\section{Theoretical description of time-resolved electron energy loss spectroscopy}
Here we give details on the theoretical description of the time-resolved electron energy loss spectroscopy (EELS). We will combine the equilibrium analysis of the EELS cross section as described in Ref.~\onlinecite{kogar2014} and a generalization of the photoemission spectroscopy to the non-equilibrium case~\cite{freericks09,eckstein2008}. In an EELS experiment the sample is probed with 
a 
pulse of electrons with definite wave-vector and energy $\ket{k_1,E_1}$, which is scattered to some final state $\ket{k_2,E_2}$. The resolved signal is proportional to the total number of electrons per solid angle $d\Omega_{k_2}$ and energy interval $dE_2$,
\beq{
  I_{k_1}(k_2)=\frac{dN(k_2)}{d\Omega_{k_2}dE_2},
}
that are emitted from the sample.

The initial state at some early time $t_{i}$ is given by the thermal ensemble of the many-body states $\ket{\Phi_n}$ in the crystal at the temperature $T$ with the density matrix $\rho(t_i) = \mathcal{Z}^{-1}\sum_n \exp[-\mathcal{E}_n(t_{i})/T] \ket{\Phi_n} \otimes \ket{k_1, E_1} \bra{k_1, E_1} \otimes \bra{\Phi_n},$ where the $\mathcal{E}_n$ are the energy eigenvalues, and the free probe electrons are denoted by $\ket{k_1,E_1}$.
This initial ensemble
$\rho(t_i)$
is evolved to some later time $t$, $\rho(t) = U(t, t_i)\rho(t_i) U(t_i, t)$, via the unitary time evolution operator $U(t,t')=\mathcal{T}\exp[-i \int_{t'}^t d\bar t (H(\bar t)+H_{\text{probe}}(\bar t))],$ where $\mathcal{T}$ is the time ordering operator. The lattice system, including the pump pulse, is described by the $H(t)$ given in Eq.~(1) of the main text and the probe electrons interact with the solid via a density-density interaction with transfers of momenta $q=k_2-k_1$ and energy $\omega = E_{2}-E_{1}$,
\begin{align}
 H_{\text{probe}} = \sum_{k,k_1,q} s(t) e^{-i\omega t} M_q(k_1) c_{k-q}^{\dagger} c_k b_{k_1+q}^{\dagger} b_{k_1},
\end{align}
where $b_{k}$ ($b^\dagger_{k}$) annihilates (creates) a probe electron with energy $E_k$ and momentum $k$, $s(t)$ is the envelope of the probe pulse at the sample (peaked around $t_p$) and $M_q(k_1)$ is the matrix element for the scattering. In the case of EELS the matrix element is proportional to the Coulomb interaction $V_q$ within the plane $M_q(k_1)\propto V_q$, see the discussion in Ref.~\onlinecite{kogar2014}. In this treatment the exchange interaction between the probe electrons and electrons in the solid is neglected as well as the presence of the surface, namely we assume that the probe-electrons measure bulk-properties of the system and that matrix elements satisfy the conservation of momentum in the plane. These simplifications can be lifted, and do not alter the general structure of the time resolved EELS response, see Ref.~\onlinecite{kogar2014} for a thorough discussion. 

The number of the detected electrons after the scattering at time $t_f$ is given by
\beq{
  N(k_2, t_f)= \Tr[ n^b_{k_2} \rho(t_f) ]
}
and the leading contribution to the measured number of electrons is given by the second-order time-dependent perturbation theory in the probe Hamiltonian $H_{\text{probe}}(t)$~\cite{freericks09} 
\bsplit{
  &N(k_2, t_f)=\iint_{t_i}^{t_f} dt dt' \Tr[ U_0(t_i,t') H_{\text{probe}}(t')\\
  &\times U_0(t',t_f) n^b_{k_2} U_0(t_f,t) H_{\text{probe}}(t) U_0(t, t_i) \rho(t_i) ],
}
where $U_0(t, t')$ is the time evolution operator of the system without probe-system coupling, $U_0(t, t') = \mathcal{T} \exp[ -i \int_{t'}^t d\bar{t} H(\bar{t}) ]$.
The evaluation of the expectation value leads to the expression for the time-resolved EELS signal, 
\begin{align}
&I_{k_1}(k_2= k_1 + q)
\propto\nonumber
|M_{q}(k_1)|^2  I_{q,\omega}\\
&I_{q,\omega}=\iint_{-\infty}^{\infty} dt dt' s(t)s(t')
e^{i\omega (t-t')}
i\chi^<_{q}(t,t'),
\label{eelstt}
\end{align}
where we sent the initial and final time to plus/minus infinity without lack of generality, $t_f,t_i \rightarrow +\infty,-\infty$, and  $\chi_{q}(t,t')
=
-i
\langle 
T_{\mathcal{C}}
n_{q}(t)
n_{- q}(t')
\rangle
$
is the momentum-dependent density-density response function of the sample 
(whose lesser component is defined as
$i\chi_{q}^<(t',t)
=
\langle 
n_{- q}(t)
n_{q}(t')
\rangle
$).
We note in passing that Eq.~\eqref{eelstt} is positive definite by construction. 

First we will show that in equilibrium Eq.~\eqref{eelstt} reduces to the conventional expression for the EELS cross-section in terms of $\text{Im}[\chi^{R}(\omega)]$~\cite{kogar2014,ayral2013}. The response of the equilibrium state is time translation invariant, $\chi_{q}^<(t',t)=\chi_{q}^<(t'-t)$ so when assuming a long probe electron envelope ($s(t)=$ const.), Eq.~\eqref{eelstt} reduces to $i\chi^<_{q}(-\omega)$. The lesser component can be related to the retarded component through the fluctuation dissipation theorem, $i\chi^<_{q}(\omega) = -2 f_B(\omega)\text{Im}[\chi^{R}_{q}(\omega)]$, where $f_B(\omega)$ is the Bose distribution function. The retarded component is odd with respect to frequency $\text{Im}[\chi^{R}_{q}(\omega)]=-\text{Im}[\chi^{R}_{q}(-\omega)]$. Thus, for frequencies larger than the temperature $| \omega | \gg 1/\beta$, where the Bose distribution function can be approximated with a Heaviside function $f_B(\omega) \approx -\theta(-\omega)$ , we have 
\begin{align}
I_{q,\omega}
\propto 
 i\chi^<_{q}(-\omega)
\approx
-2\theta(\omega)\text{Im}[\chi^{R}_{q}(\omega)].
\end{align}
In other words, in this limit there is only energy loss ($\omega > 0$), and the EELS cross-section is given by the spectral density of the valence electron density-density response function.

Out of equilibrium, the convolution with the probe envelopes $s(t)$ can be viewed as a filter for the susceptibility $\chi_{q}^<(t,t')$ in the time-frequency plane, similar to the case of photoemission spectroscopy \cite{randi2017}. In the simplest approximation, we assume a Gaussian wave packet $s(t)\propto e^{-{(t-t_p)^2}/{2\delta t^2}}$ of duration $\delta t$ centered around a probe time $t_p$. Then Eq.~\eqref{eelstt} becomes the convolution of the Wigner transform 
\begin{align*}
\chi^<_{q}(t,\omega)=
\int_{-\infty}^\infty ds \, e^{i\omega s} \chi^<_{q}(t+s/2,t-s/2)
\end{align*}
with a Gaussian kernel of width $\delta t $ and $\delta \omega=1/\delta t$ centered at frequency $-\omega$ and time $t_p$. As in the case of photoemission spectroscopy, if the evolution of the spectrum is fast compared to the inverse width of relevant spectral signatures, the form of the spectrum will strongly depend on the time-profile $s(t)$ of the probe pulse \cite{randi2017}. In the present paper, however, we characterize the spectrum of the system on the frequency scale of the order of the bandwidth, in a non-thermal steady state that lives much longer than the inverse bandwidth. In such a non-thermal quasi-steady state we can approximate $I_{q,\omega}$ by the Wigner transform \footnote{In fact, any Fourier transform with respect to a time-difference is equivalent in a steady state, up to a phase factor.},
\begin{align}
I_{q,\omega}
\propto
i\chi^<_{q}(t_p,-\omega).
\label{quasi0steady}
\end{align}
This is the approximation for the EELS cross-section employed in this paper.

In general, there will be both energy loss and gain, as described by the signal $I_{q,\omega}$ at positive and negative frequencies, respectively. In both cases, the result given by Eq.~\eqref{eelstt} is positive definite. To detect the population inversion experimentally, one can evaluate the \emph{difference} $\Delta I_{q,\omega}$ between the gain and the loss,
\begin{multline}
\Delta I_{q,\omega}
=
I_{q,\omega}-I_{q,-\omega}
\propto
\iint_{-\infty}^\infty dt dt' \,
s(t)s(t') \\
 \times \Big[
e^{i\omega (t-t')}
\langle
n_{-q}(t)
n_{q}(t')
\rangle
-
e^{-i\omega (t-t')}
\langle
n_{-q}(t)
n_{q}(t')
\rangle
\Big]
\\
=
\iint_{-\infty}^\infty dt dt' \,
s(t)s(t')
e^{i\omega (t-t')}
i[\chi_{q}^<(t',t)
-
\chi_{q}^>(t',t)
]
\\
=
2\text{Im}
\iint_{-\infty}^\infty dt dt'
s(t)s(t')
e^{i\omega (t-t')}
\chi^{R}_{q}(t',t),
\label{eq:Diff}
\end{multline}
defined for positive frequencies $\omega > 0$. In the third line of Eq.~\eqref{eq:Diff} we have relabeled the time arguments $t\leftrightarrow t'$ in the second term under the integral, and assumed momentum inversion symmetry $\chi_{q}(t,t')=\chi_{-q}(t,t')$], while in the fourth line we have used the anti-hermiticity relation $\chi_{q}^{>,<}(t,t')=-\chi_{q}^{>,<}(t',t)^*$. In the quasi-steady state we employ the same approximation as in Eq.~\eqref{quasi0steady}, and obtain the difference between the EELS energy-gain and energy-loss cross section as
\begin{align}
\Delta I_{q,\omega}
\approx
2\text{Im}\chi^{R}_{q}(t,-\omega) =
-2\text{Im}\chi^{R}_{q}(t,\omega),
\end{align}
which lead to the EELS signal
\beq{
  I_{k_1}(k_2=k_1+q)\propto V_{q}^2 \text{Im} \chi^R_{q}(t,\omega).
}
Hence the {\em difference in energy gain and energy loss} at energy $\omega$ can be used to detect the transient population-inversion in the density fluctuations of the system. 
In Fig.~4 of the main text we plot the related quantity  $V_{q} \text{Im} \chi^R_{q}(t,\omega)$, which corresponds to the inverse dielectric constant.